\documentclass[showpacs,amsmath,amssymb,twocolumn,prx,longbibliography,superscriptaddress,preprintnumbers,nofootinbib,superscriptaddress]{revtex4-2}
\usepackage{amssymb}
\usepackage[dvips]{graphicx}
\usepackage{enumerate}
\usepackage{epsfig}
\usepackage{subfigure}
\usepackage{xcolor}
\usepackage[T1]{fontenc}
\usepackage{fullpage}
\usepackage{amsthm,amsfonts,amssymb,amscd,mathrsfs,xspace,framed}
\usepackage{threeparttable}
\usepackage{mathrsfs,amsmath}
\usepackage[linguistics]{forest}
\usepackage{color}
\usepackage{enumitem}
\usepackage{setspace}
\usepackage{url}
\usepackage{eqparbox} 
\newcommand{\okina}{\textquotesingle}
\usepackage{wrapfig}
\usepackage{array,multirow}
\usepackage{enumitem}
\usepackage{bm}
\usepackage{comment}
\usepackage{txfonts}
\usepackage{array}
\usepackage{upgreek}
\newtheoremstyle{nonitalic}
  {3pt}   % Space above
  {3pt}   % Space below
  {\normalfont}  % Body font
  {}      % Indent amount
  {\bfseries}  % Theorem head font
  {.}     % Punctuation after theorem head
  { }     % Space after theorem head
  {}      % Theorem head spec (can be left empty, meaning `normal`)

\theoremstyle{nonitalic}

\usepackage{braket}
\usepackage{mathtools}

\usepackage[all]{xy}

\usepackage[hyperfootnotes=false]{hyperref}
\usepackage[acronym]{glossaries}
\usepackage[english]{babel}
\usepackage{url}
\usepackage{hyperref}
\definecolor{darkblue}{rgb}{0,0,0.5}
\hypersetup{
colorlinks=true,
linkcolor=black,
filecolor=blue,
citecolor=darkblue,  
urlcolor=black,
}
\usepackage{gensymb}

\usepackage{xcolor}

\newcommand{\BH}[1]{{{\textcolor{black}{#1}}}}

\begin{document}

\title{Breaking On/Off-coupling Loss Degeneracies via Bidirectional Nonlinear Optics}

\author{Bo-Han Wu}
\email{bohanwu@hawaii.edu}
\affiliation{Research Laboratory of Electronics, Massachusetts Institute of Technology, Cambridge, Massachusetts 02139, USA}
\affiliation{Electrical and Computer Engineering, University of Hawai\okina i at M\={a}noa, Honolulu, Hawai\okina i 96822, USA}

\author{Mahmoud Jalali Mehrabad}
\affiliation{Research Laboratory of Electronics, Massachusetts Institute of Technology, Cambridge, Massachusetts 02139, USA}

% \author{Xinyi Ren}
% \affiliation{Department of Electrical Engineering and Computer Sciences, University of California, Berkeley, CA 94720, USA}

\author{Mengjie Yu}
\affiliation{Department of Electrical Engineering and Computer Sciences, University of California, Berkeley, CA 94720, USA}

\author{Dirk Englund}
\affiliation{Research Laboratory of Electronics, Massachusetts Institute of Technology, Cambridge, Massachusetts 02139, USA}

\date{\today}

%%%%%%%%%%%%%%%%%%%
\begin{abstract}
Accurate characterization of nonlinear photonic integrated circuits is fundamentally limited by the inability of linear transmission measurements to distinguish input and output \BH{on/off-chip coupling} efficiencies, which enter only through their product. We show that this degeneracy arises from an inherent input–output symmetry of linear optics and can be broken by exploiting the directional asymmetry of nonlinear processes. We introduce bidirectional nonlinear optical tomography (BNOT), which combines forward and backward pumping of complementary nonlinear interactions to uniquely infer individual \BH{on/off-coupling} efficiencies from off-chip measurements. By jointly modeling the nonlinear response and detection statistics, BNOT enables unbiased reconstruction of on-chip nonlinear figures of merit with significantly reduced uncertainty, whereas conventional calibration yields systematically biased estimates. Monte Carlo simulations demonstrate reliable convergence of the inferred efficiencies across realistic noise and fluctuation regimes. Our results establish nonlinear directionality as a general symmetry-breaking resource for coupling-resolved characterization in integrated photonics, with immediate implications for scalable quantum light sources, frequency conversion, and precision optical metrology.
\end{abstract}

\maketitle
\begin{figure*}
    \centering
	{\centering\includegraphics[width=0.88\linewidth]{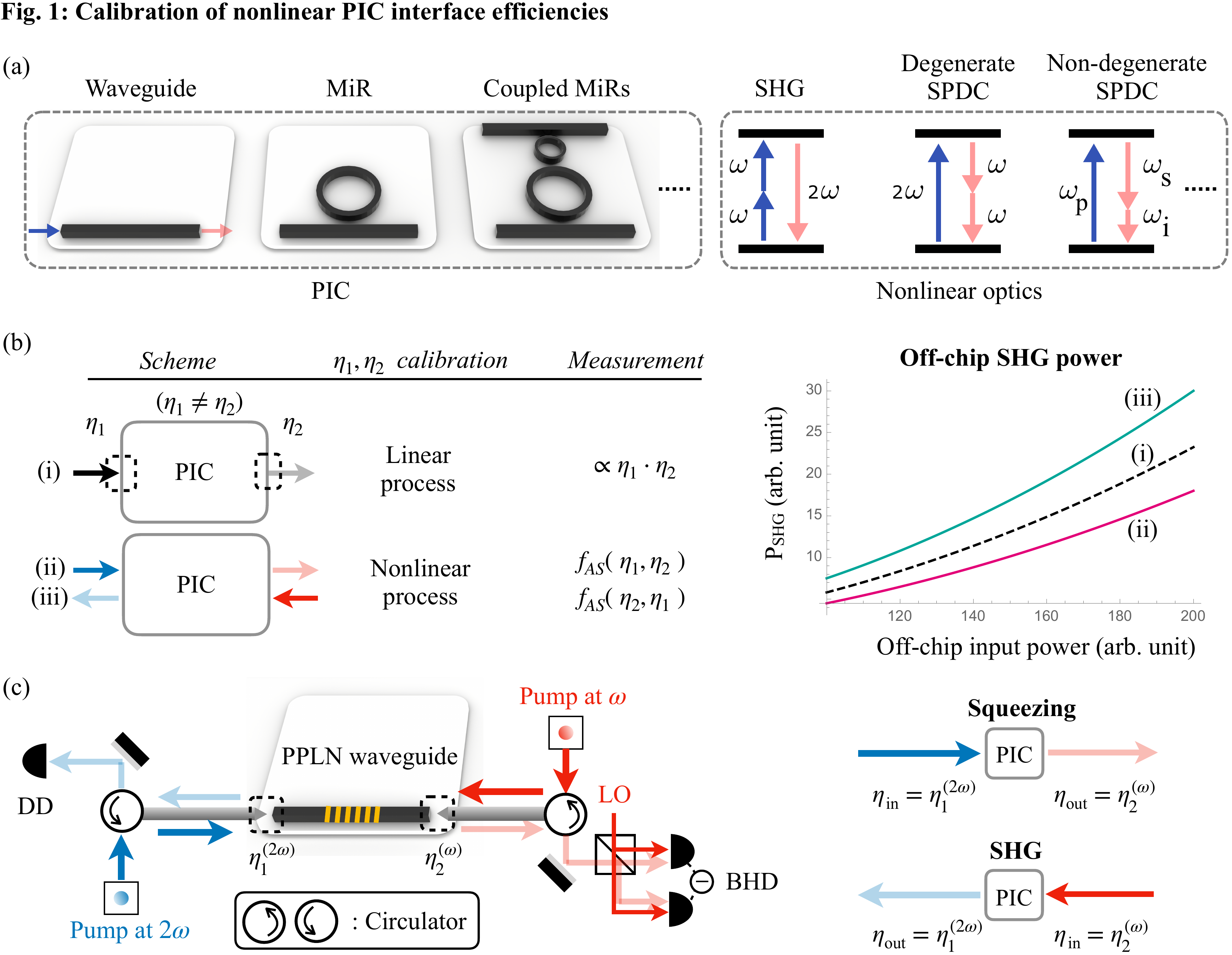}}
	\caption{(a) Representative integrated photonic structures and nonlinear optical processes. (b) Calibration protocols. Linear transmission measurements yield a response proportional to $\eta_1\cdot\eta_2$ and are invariant under exchange of the input and output interfaces. In contrast, nonlinear processes probed under forward (left-to-right) and backward (right-to-left) pumping exhibit asymmetric responses described by $f_{AS}(\eta_1,\eta_2)$ and $f_{AS}(\eta_2,\eta_1)$. The right panel shows the corresponding scaling of second-harmonic generation (SHG) with input power for each scheme. (c) Bidirectional nonlinear optical tomography (BNOT) implemented in a PPLN waveguide. Pumps at $\omega$ and $2\omega$ generate SHG and squeezed light under forward pumping ($\eta_\text{in}=\eta^{(2\omega)}_1$, $\eta_\text{out}=\eta^{(\omega)}_2$) and backward pumping ($\eta_\text{in}=\eta^{(\omega)}_2$, $\eta_\text{out}=\eta^{(2\omega)}_1$). The output fields are measured using direct detection (DD) and balanced homodyne detection (BHD) with a local oscillator (LO).}
	\label{fig:Scheme}
\end{figure*}
\textit{Introduction}---Nonlinear photonic integrated circuits (PICs) promise a compact wafer-scale platform that transforms passive chips into active nonlinear engines for frequency conversion and generation of quantum states~\cite{boyd2008nonlinear,Leuthold2010}. Across emerging platforms such as lithium niobate (LiN), silicon nitride (SiN), aluminum nitride, and gallium arsenide, canonical processes, including spontaneous parametric down conversion (SPDC)~\cite{nehra2022few,williams2024ultrashort,chen2022ultra}, four-wave mixing (FWM)~\cite{shen2025strong,aghaee2025scaling,pang2025versatile,shen2025highly,dutt2015chip,yang2021squeezed,liu2025wafer,liu2025ultra}, harmonic generation~\cite{lu2021efficient,mehrabad2025multi,dean2025low,yang2024symmetric}, self-phase modulation (SPM)~\cite{cernansky2020nanophotonic}, optical parametric oscillation (OPO)~\cite{ji2017ultra,briles2020generating,flower2024observation,xu2025chip}, optical parametric amplification (OPA)~\cite{kashiwazaki2023over,nehra2022few,dean2025low} and \BH{microcomb generation~\cite{song2025stable,chang2020ultra}}, are now accessible on chip, enabling applications in spectroscopy, sensing, and quantum information science~\cite{nehra2022few,shen2025strong,aghaee2025scaling,pang2025versatile,shen2025highly,lu2021efficient,mehrabad2025multi,ji2017ultra,briles2020generating,Wang2019Monolithic,song2025stable,chang2020ultra}. These processes are carried out in diverse PIC structures (Fig.~\ref{fig:Scheme}(a)), ranging from waveguides to a variety of microring resonators, which provide tailored dispersion and mode confinement. Multimode and microcomb architectures further scale brightness and mode count with favorable hardware efficiency~\cite{guidry2023multimode}, positioning nonlinear PICs as strong candidates for continuous-variable (CV) quantum computing~\cite{aghaee2025scaling,fukui2018high,wu2020quantum} and classical optical computing~\cite{feldmann2019all,mesaritakis2013micro,bandyopadhyay2021hardware,wu2025micro}.

Despite these advances, a fundamental obstacle to quantitative benchmarking of nonlinear PICs remains: the inability to uniquely determine on/off chip coupling efficiencies from standard measurements. Because nonlinear experiments rely on off-chip lasers and detectors, observed signals depend on two \BH{on/off-coupling} efficiencies $(\eta_1,\eta_2)$ that are generally unequal due to mode mismatch, polarization dependence, fabrication imperfections, or asymmetric coupling schemes. Linear transmission measurements are invariant under the exchange $\eta_1 \leftrightarrow \eta_2$ and therefore recover only their product $\eta_1\cdot\eta_2$, enforcing a ``degenerate'' estimate $(\eta_1\cdot\eta_2)^{1/2}$ that systematically biases inferred on-chip performance~\cite{Lomonte2021,Hansen2023,sacher2014}. This limitation becomes explicit in nonlinear processes such as second-harmonic generation (SHG), where the detected off-chip efficiency scales asymmetrically as $P_\text{SHG}/P_\text{in}^2 \propto \eta_\text{in}^2\,\eta_\text{out}$~\cite{Yakar2023PRL_BackwardSHG}. Forward and backward pumping thus yield distinct responses, $f_\text{AS}(\eta_1,\eta_2)\neq f_\text{AS}(\eta_2,\eta_1)$ (Fig.~\ref{fig:Scheme}(b)), whereas symmetric calibration incorrectly enforces $\eta_\text{in}=\eta_\text{out}$. While bidirectional SHG has been used to qualitatively attribute performance differences to asymmetric coupling~\cite{wang2018ultrahigh}, existing approaches remain unable to uniquely resolve individual \BH{on/off-coupling} efficiencies. This symmetry-induced degeneracy fundamentally limits bias-free inference of on-chip nonlinear performance.
\begin{figure*}
    \centering
	{\centering\includegraphics[width=1\linewidth]{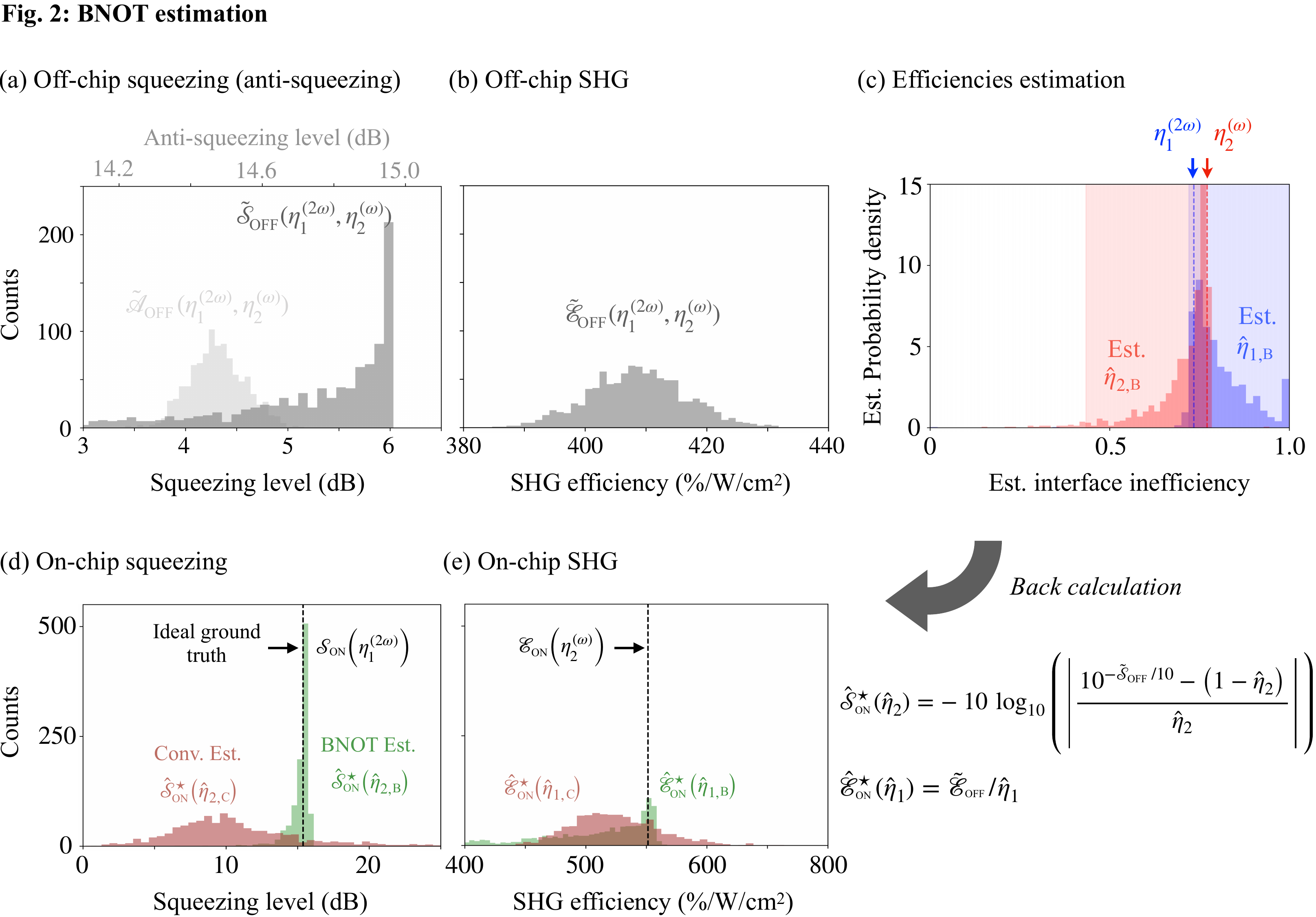}}
	\caption{\BH{(a) Monte Carlo (MC) histogram of simulated off-chip squeezing. (b) MC histogram of simulated off-chip SHG efficiency. (c) BNOT estimates of the \BH{on/off-coupling} efficiencies $\hat{\eta}_{1,\text{B}}$ (blue) and $\hat{\eta}_{2,\text{B}}$ (red). Shaded regions indicate the 95\% confidence intervals (2.5–97.5 percentiles), and dashed vertical lines mark the ground-truth efficiencies $\eta^{(2\omega)}_1$ and $\eta^{(\omega)}_2$. (d) Histogram of the inferred on-chip squeezing based on BNOT (green) and the conventional calibration approach (brown). (e) Histogram of the inferred on-chip SHG efficiency using the same procedures. The BNOT on-chip quantities are obtained by back-calculating the simulated off-chip observables using the estimated efficiencies $\hat{\eta}_{1,\text{B}}$ and $\hat{\eta}_{2,\text{B}}$ from (c). Black dashed lines in (d,e) denote the ideal ground-truth values $\mathcal{S}_\text{ON}(\eta^{(2\omega)}_1)$ and $\mathcal{E}_\text{ON}(\eta^{(\omega)}_2)$.}}
	\label{fig:Histogram}
\end{figure*}

Among nonlinear processes, squeezed light provides another stringent benchmark for on-chip device performance. Bulk-optics squeezing experiments have reached $15~\text{dB}$~\cite{vahlbruch2016detection}, while envisioned applications require near-$10~\text{dB}$ for CV fault-tolerant quantum computing~\cite{fukui2018high}, gravitational-wave detectors~\cite{aasi2013enhanced}, and Gottesman-Kitaev-Preskill (GKP) sources~\cite{larsen2025nature}. Integrated platforms have advanced from $1.7~\text{dB}$ demonstrations~\cite{dutt2015chip} to multi-decibel squeezing in LiN~\cite{aghaee2025scaling}, cavity optomechanics~\cite{safavi2013squeezed}, and foundry-compatible SiN~\cite{dutt2016tunable,liu2025wafer}, with further gains in nanophotonic molecules and microcombs~\cite{zhang2021squeezed,yang2021squeezed}. The measured squeezing levels reported (off-chip) now exceed $4.9~\text{dB}$ in PPLN waveguides, $3.5~\text{dB}$ in Kerr microrings, $5.6~\text{dB}$ in microcombs~\cite{nehra2022few,shen2025strong,shen2025highly}, and attain $3.1~\text{dB}$ in wafer-scale integration~\cite{liu2025wafer}. As measured values increase and target chip squeezing thresholds approach $10-15~\text{dB}$, an accurate estimation of \BH{on/off-coupling} efficiencies $(\eta_1,\eta_2)$ becomes even more critical~\cite{Park2024,Zhang2021,shen2025strong,zhao2020high,chen2026universal}.

We introduce \emph{bidirectional nonlinear optical tomography} (BNOT), a directionally-sensitive metrology that combines forward- and backward-pumped nonlinear probes to break the $\eta_1\cdot\eta_2$ ``degeneracy'' inherent to linear transmission. BNOT enables direct estimation of the individual on/off chip coupling efficiencies with confidence intervals that are significantly narrower than those obtained from conventional linear calibration. \BH{Recent efforts have explored advanced characterization strategies for nonlinear photonic systems~\cite{goh2026visible}; however, resolving the intrinsic degeneracy between input and output coupling efficiencies remains an open challenge.}

\textit{Concept}---Our BNOT methodology leverages the reversibility of two nonlinear processes within the same photonic device by pumping it in opposite directions. To benchmark its performance, we consider on-chip squeezed-light generation and SHG in an exemplary platform: periodically poled LiN (PPLN) waveguides. The calibration scheme is illustrated in Fig.~\ref{fig:Scheme}(c). A pump field at $2\omega$ is injected into the PPLN waveguide through the \textit{left} interface in the forward direction (\textit{left-to-right}), while a pump at $\omega$ is simultaneously launched through the \textit{right} interface in the backward direction (\textit{right-to-left}). The corresponding on/off chip coupling efficiencies are denoted $\eta_1^{(2\omega)}$ and $\eta_2^{(\omega)}$, where the superscript denotes the pump frequency. This bidirectional pumping exploits the reversibility of the nonlinear interactions, enabling both degenerate SPDC and SHG to occur concurrently. The corresponding output fields are measured at their respective ports using balanced homodyne detection and direct detection.

To relate on- and off-chip performances in squeezing and SHG, we define two estimation variables, $x_1$ and $x_2$, corresponding to the \textit{left}- and \textit{right}-coupling efficiencies of the PPLN waveguide. In particular, these are not the ground-truth efficiencies: $\eta_1^{(2\omega)}$ and $\eta_2^{(\omega)}$ but the tunable parameters in our model used to recover them. The off-chip anti-squeezing and squeezing levels (in dB), and SHG efficiency (in W$^{-1}$m$^{-2}$) are expressed as
\BH{\begin{equation}
\begin{aligned}
        \mathcal{A}_{\text{OFF}}\,(x_1,x_2)&=10\log_{10}\left(x_2\,10^{\mathcal{A}_{\text{ON}}\,(x_1)/10}+1-x_2\right),\\
        \mathcal{S}_{\text{OFF}}\,(x_1,x_2)&=-10\log_{10}\left(x_2\,10^{-\mathcal{S}_{\text{ON}}\,(x_1)/10}+1-x_2\right),\\
        \mathcal{E}_{\text{OFF}}\,(x_1,x_2)&=x_1\,\mathcal{E}_{\text{ON}}(x_2),
        \label{eq:Sqz_SHG}
\end{aligned}
\end{equation}
where $\mathcal{S}_\text{ON}$ and $\mathcal{E}_\text{ON}$ denote the analytical models for the on-chip squeezing~\cite{kashiwazaki2020continuous} and SHG efficiency~\cite{yang2024symmetric}, respectively, with their derivations provided in App.~\ref{sec:analytical}.}

Conventional calibration approaches based on linear transmission cannot distinguish between \textit{left-to-right} and \textit{right-to-left} propagation~\cite{Lomonte2021,Hansen2023,sacher2014,Lomonte2021}. As a result, they impose a ``degenerate’’ interface assumption, leading to systematic estimation bias. \BH{In practice, the estimation becomes even more challenging because experimental uncertainties propagate through the analysis, particularly those associated with the following types of fluctuations: a) pump-power fluctuations, b) temperature-induced phase-matching drift, c) phase noise in homodyne detection.}
\begin{figure*}
    \centering
	{\centering\includegraphics[width=0.9\linewidth]{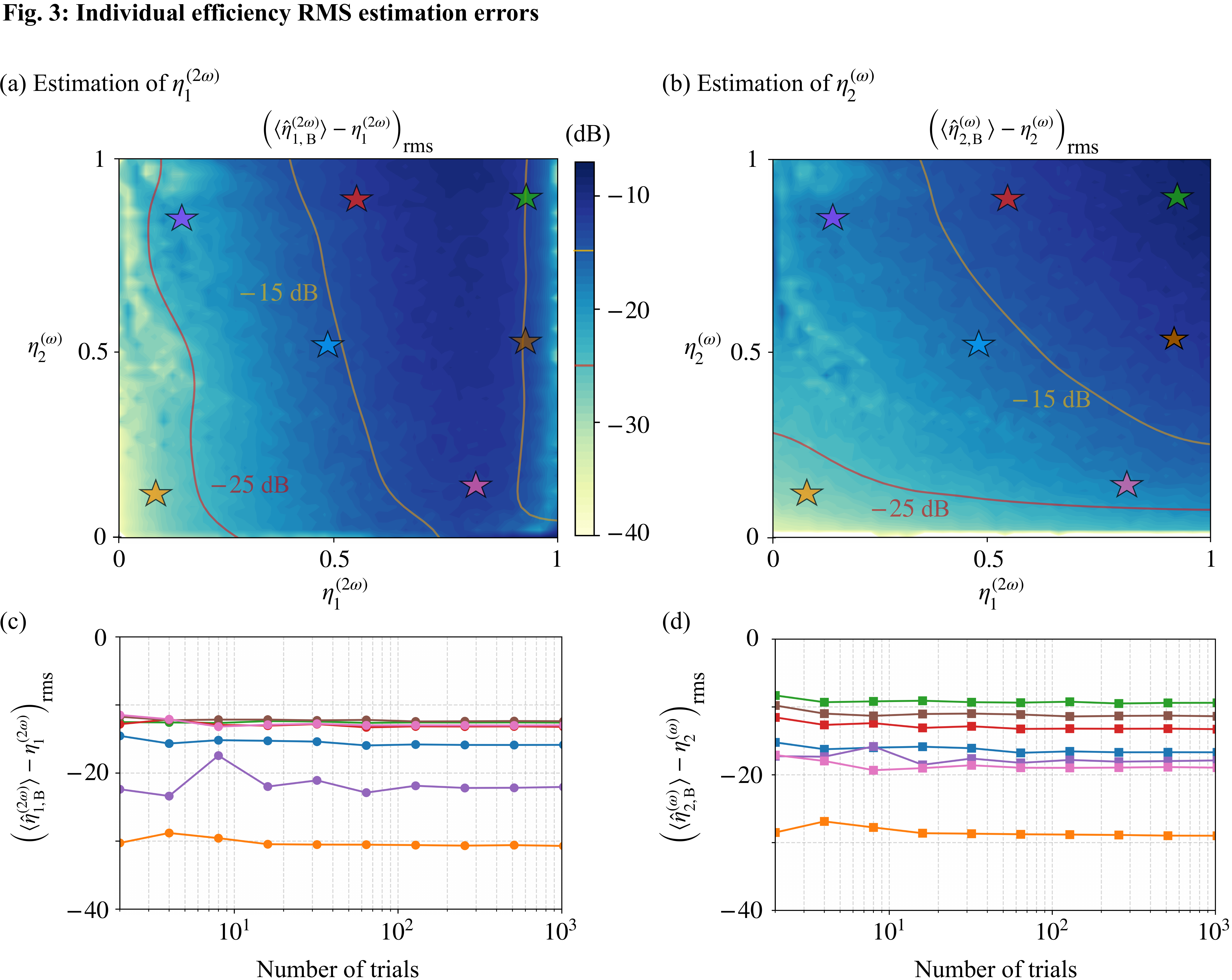}}
	\caption{\BH{(a) Root-mean-square (RMS) estimation error of $\eta_1^{(2\omega)}$ obtained using BNOT. (b) RMS estimation error of $\eta_2^{(\omega)}$ obtained using BNOT. The red and orange contours in (a,b) indicate the $-25$~dB and $-15$~dB error boundaries. (c) RMS error of $\eta_1^{(2\omega)}$ as a function of the number of MC trials for seven representative pairs of $\left\{\eta_1^{(2\omega)},\eta_2^{(\omega)}\right\}$. (d) RMS error of $\eta_2^{(\omega)}$ versus the number of MC trials for the same representative pairs.}}
	\label{fig:MSE}
\end{figure*}

To overcome this challenge, we adopt an optimization-based estimation approach that systematically searches the parameter space to obtain the optimal estimates $\hat{\eta}_{1,\text{B}}$ and $\hat{\eta}_{2,\text{B}}$ from the simulated off-chip nonlinear optical observables.
\\

\textit{Simulations}---Using the ground truth efficiencies 
\begin{equation}
\begin{aligned}
\eta_1^{(2\omega)}&=0.734,\quad\eta_2^{(2\omega)}=0.794,\quad\eta_2^{(\omega)}=0.771,
\end{aligned}
\label{eq:ground_truth}
\end{equation}
together with the physical parameters summarized in Tab.~\ref{tab:parameters}, \BH{Figs.~\ref{fig:Histogram}(a,b) shows Monte Carlo (MC) histograms of the simulated off-chip observables: squeezing $\tilde{\mathcal{S}}_\text{OFF}(\eta_1^{(2\omega)},\eta_2^{(\omega)})$, anti-squeezing $\tilde{\mathcal{A}}_\text{OFF}(\eta_1^{(2\omega)},\eta_2^{(\omega)})$, and SHG efficiency $\tilde{\mathcal{E}}_\text{OFF}(\eta_1^{(2\omega)},\eta_2^{(\omega)})$. The tilde ``$\sim$’’ indicates that the observables incorporate the three experimental fluctuations discussed above: pump power fluctuations, wave-vector mismatch, and local-oscillator phase noise, thereby emulating realistic measurement data.}

Utilizing the previously defined models in Eq.~\eqref{eq:Sqz_SHG}, we exhaustively search for the tunable efficiencies $x_1$ and $x_2$ such that the weighted estimation errors, $e^2_\text{sq}(x_1,x_2)$, $e^2_\text{asq}(x_1,x_2)$ and $e^2_\text{shg}(x_1,x_2)$, from the three models can be minimized:
\BH{\begin{widetext}
    \begin{equation}
    \begin{aligned}
            \hat{\eta}_{1,\,\text{B}}\;,\; \hat{\eta}_{2,\,\text{B}} &= \underset{\{x_1,\,x_2,\,w\}}{\text{argmin}}
            \left\{e_\text{sq}^2(x_1,x_2)+e_\text{asq}^2(x_1,x_2)+w\,e_\text{shg}^2(x_1,x_2)
            \right\},\\\\
            e^2_\text{sq}(x_1,x_2)\equiv\left|1-\frac{\mathcal{S}_\text{OFF}(x_1,x_2)}{\tilde{\mathcal{S}}_\text{OFF}(\eta^{(2\omega)}_1,\,\eta^{(\omega)}_2)}\right|^2\;,\; &e^2_\text{asq}(x_1,x_2)\equiv\left|1-\frac{\mathcal{A}_\text{OFF}(x_1,x_2)}{\tilde{\mathcal{A}}_\text{OFF}(\eta^{(2\omega)}_1,\,\eta^{(\omega)}_2)}\right|^2\;,\; e^2_\text{shg}(x_1,x_2)\equiv\left|1-\frac{\mathcal{E}_\text{OFF}(x_1,x_2)}{\tilde{\mathcal{E}}_\text{OFF}(\eta^{(2\omega)}_1,\,\eta^{(\omega)}_2)}\right|^2\\
    \end{aligned}
    \label{eq:optimization}
\end{equation}
\end{widetext}
i.e., $x_1\,,\,x_2\in(0,1)$ and $w\in\mathbb{R}^+$. Here, $w$ controls the relative weighting between the squeezing (and anti-squeezing) error and the SHG error.} In words, the \BH{on/off-coupling} efficiencies $\eta^{(2\omega)}_1$ and $\eta^{(\omega)}_2$ are inferred by jointly fitting the squeezing, anti-squeezing and SHG observables, yielding a pair of estimates $\hat{\eta}_{1,\text{B}}$ and $\hat{\eta}_{2,\text{B}}$ that is simultaneously consistent with both nonlinear responses.

\begin{table}[h]
    \begin{tabular}{ccc}
        \multicolumn{3}{c}{\textbf{Parameters summary}} \\
         \hline\hline\\
         $d_{33}$\;&\;$19.5$~pm/V\;&Thin-film LiN \\
         &\cite{yang2024symmetric}&nonlinear coefficient\\\\
         $L$\;&\;\BH{$5$~mm}\;&\;Poling length of PPLN\\\\
     $\omega$\;&$193$~THz\;& Pump frequency\\
     (or $2\omega$)\;&\;(or $386$~THz)\;\;&\;\;at $1550$~nm (or $775$~nm)\\\\
         $n_{\omega}$\;&$2.14$\;&\;Refractive index of LiN \\
         (or $n_{2\omega}$)\;&(or $2.2$)\;&\;at $\omega$ (or $2\omega$)\\\\
         \BH{$R_{\omega}$}\;&\BH{$0.13$}\;&\;\BH{Reflectivity of waveguide's}\\
         \BH{(or $R_{2\omega}$)}\;&\BH{(or $0.14$)}\;&\;\BH{facet at $\omega$ (or at $2\omega$)}\\\\
         $\mathcal{S}_{\omega}$\;&\;$1$~\textmu m$^2$\;& Mode area in PPLN\\
         (or $\mathcal{S}_{2\omega}$)&\;(or $0.8$~\textmu m$^2$)\;& waveguide\\\\
         $\Gamma_{\omega}$&$5$~dB/m~\cite{kashiwazaki2021fabrication}&Propagation loss\\
         (or $\Gamma_{2\omega}$)&(or $10$~dB/m)&\\\\
         $\sigma_\text{PN}$&$0.05$~rad\;&\;LO phase noise in\\
         &&\;squeezing experiment\\\\
         $P_{2\omega}$\;&\BH{$200$~mW} &Average pump power\\
         && in squeezing experiment\\\\
         $\sigma_{2\omega}$&$0.5~\%$&Pump fluctuation rate\\
         (or $\sigma_\omega$)&&in squeezing (or SHG)\\
         & & \\
         $\sigma_{\Delta k}$& \BH{$0.1$~m$^{-1}$}& Wave-vector mismatch\\
         &&STD\\\\
         $\sigma_\text{C}$&$7~\%$& Linear transmission\\
         && measurement STD\\\\
         \hline\hline
    \end{tabular}
    \caption{Simulation parameters. STD: standard deviation.}
    \label{tab:parameters}
\end{table}

The effectiveness of this joint inference arises from the asymmetric physical responses encoded in the three nonlinear observables. Propagation loss inside the waveguide mixes the intracavity field with vacuum noise, which affects the squeezed and anti-squeezed quadratures with different magnitudes and therefore provides distinct sensitivities to the \BH{on/off-coupling} efficiencies. In addition, the measurement noise mechanisms differ between the observables: squeezing and anti-squeezing are measured using balanced homodyne detection and are therefore affected by local-oscillator phase noise, whereas the SHG observable is an intensity measurement and is insensitive to phase fluctuations. These asymmetries make the three observables complementary, allowing the optimization in Eq.~\eqref{eq:optimization} to extract sufficient information to determine both \BH{on/off-coupling} efficiencies.

Fig.~\ref{fig:Histogram}(c) presents the histograms of the estimators $\hat{\eta}_{1,\text{B}}$ and $\hat{\eta}_{2,\text{B}}$, which are confirmed to be unbiased since their mean values satisfy $\langle\hat{\eta}_{1,\text{B}}\rangle \approx \eta_1^{(2\omega)}$ and $\langle\hat{\eta}_{2,\text{B}}\rangle \approx \eta_2^{(\omega)}$. On the other hand, in conventional estimation approach, the estimator is a random variable
\begin{equation}
\hat{\eta}_{1,\text{C}}=\hat{\eta}_{2,\text{C}}=\mathcal{N}\left(\sqrt{\eta_1^{(\omega)}\,\eta_2^{(\omega)}},\sigma_\text{C}^2\right).
\end{equation}
Here, $\sigma_\text{C}$ denotes the standard deviation associated with linear calibration performed at frequency $\omega$.

\BH{Based on the estimated efficiencies $\hat{\eta}_1=\{\hat{\eta}_{1,\text{B}},\hat{\eta}_{1,\text{C}}\}$ and $\hat{\eta}_2=\{\hat{\eta}_{2,\text{B}},\hat{\eta}_{2,\text{C}}\}$, we back-calculate the on-chip squeezing and SHG using the noiseless relations in Eq.~\eqref{eq:Sqz_SHG}:
\begin{equation}
\begin{aligned}
        \hat{\mathcal{S}}^\star_\text{ON}(\hat{\eta}_2)&=-10\,\log_{10}\left(\frac{10^{-\tilde{\mathcal{S}}_\text{OFF}(\eta^{(2\omega)}_1,\,\eta^{(\omega)}_2)/10}-\left(1-\hat{\eta}_2\right)}{\hat{\eta}_2}\right)\\
        \hat{\mathcal{E}}^\star_\text{ON}(\hat{\eta}_1)&=\tilde{\mathcal{E}}_\text{OFF}(\eta^{(2\omega)}_1,\,\eta^{(\omega)}_2)/\hat{\eta}_1.
        \label{eq:estimation_on_chip}
\end{aligned}
\end{equation}
Figs.~\ref{fig:Histogram}(d,e) show the estimated on-chip performance, where the black dashed vertical lines indicate the ideal ground-truth values $\mathcal{S}_\text{ON}(\eta^{(2\omega)}_1)$ and $\mathcal{E}_\text{ON}(\eta^{(\omega)}_2)$. Unlike the estimators $\hat{\mathcal{S}}^\star_\text{ON}$ and $\hat{\mathcal{E}}^\star_\text{ON}$ in Eq.~\eqref{eq:estimation_on_chip}, the quantities $\mathcal{S}_\text{ON}(\eta^{(2\omega)}_1)$ and $\mathcal{E}_\text{ON}(\eta^{(\omega)}_2)$ do not incorporate stochastic fluctuations of the physical system and therefore represent the ideal on-chip squeezing and SHG performance.}

\BH{The expectation values of the estimated on-chip squeezing and SHG agree with these ideal quantities, shown in Figs.~\ref{fig:Histogram}(e,f), confirming the unbiasedness of the BNOT estimator. In contrast, the conventional calibration approach produces biased estimates with significantly broader distributions. This difference arises because BNOT jointly infers the efficiencies by enforcing consistency between two invertible nonlinear optical processes, thereby reducing estimation uncertainty. On the other hand, the conventional method relies on an independent linear calibration of coupling efficiencies, which introduces additional calibration uncertainty into the inferred on-chip performance.}

Additionally, within the BNOT results, the different spreads of the inferred on-chip squeezing and SHG distributions reflect the distinct information encoded in the two nonlinear observables. The SHG efficiency constrains an asymmetric combination of $\eta^{(2\omega)}_1$ and $\eta^{(\omega)}_2$ through its directional nonlinear scaling, whereas the squeezing observable is primarily sensitive to the output efficiency $\eta^{(\omega)}_2$ via vacuum-noise admixture at the detection port. Consequently, the squeezing estimates are more tightly concentrated, while the SHG estimates exhibit a broader distribution due to their joint dependence on both \BH{on/off-coupling} efficiencies. Individually, neither observable uniquely determines the two efficiencies; only their joint use breaks the intrinsic $\eta_1\cdot\eta_2$ degeneracy.

To assess the robustness of BNOT beyond the \BH{particularly assigned on/off-coupling efficiencies in Fig.~\ref{fig:Histogram}}, we evaluate the estimator across the full parameter space of ground-truth efficiencies \BH{$\eta^{(2\omega)}_{1},\eta^{(\omega)}_{2}\in(0,1)$. Figs.~\ref{fig:MSE}(a,b) show the root-mean-square (RMS) errors of the two estimated efficiencies. For each point in Fig.~\ref{fig:MSE}(a), the MC trials are partitioned into subsets to obtain multiple subset-averaged estimates of $\hat{\eta}_{1,\text{B}}$, whose deviations from the ground-truth value $\eta^{(2\omega)}_{1}$ are then used to compute the RMS error, denoted as $|\langle\hat{\eta}_{1,\text{B}}\rangle-\eta^{(2\omega)}_{1}|_\text{rms}$. The same procedure is applied in Fig.~\ref{fig:MSE}(b) for $\hat{\eta}_{2,\text{B}}$ relative to $\eta^{(\omega)}_{2}$, denoted as $|\langle\hat{\eta}_{2,\text{B}}\rangle-\eta^{(\omega)}_{2}|_\text{rms}$. The red and yellow contours indicate the $-25$~dB and $-15$~dB RMS error boundaries, respectively. Fig.~\ref{fig:Total_RMSE} in Appendix~\ref{sec:Total_RMSE} shows the combined estimation RMS error of $\eta^{(2\omega)}_{1}$ and $\eta^{(\omega)}_{2}$, showing the overall joint estimation accuracy of the two efficiencies.}

To further illustrate the convergence behavior, seven representative parameter pairs $(\eta_1^{(2\omega)},\eta_2^{(\omega)})$, indicated by the colored stars in Figs.~\ref{fig:MSE}(a,b), are examined in Figs.~\ref{fig:MSE}(c,d). These panels show that the estimation errors decrease and gradually stabilize as the number of MC trials increases.
\\

%%%%%%%%%%%%%%%%%%%
\textit{Conclusion and outlook}---In summary, bidirectional nonlinear optical tomography (BNOT) provides a platform-agnostic approach for resolving asymmetric coupling efficiencies in nonlinear photonic integrated circuits. As demonstrated using SHG and squeezed-light measurements \BH{in $\chi^{(2)}$ materials}, BNOT integrates the existing experimental platforms and enables accurate, coupling-resolved benchmarking of on/off-chip efficiencies. \BH{More broadly, BNOT reflects a general measurement principle: conjugate nonlinear processes with distinct directional sensitivities, arising from controlled asymmetries in the model. For example, in a $\chi^{(3)}$ medium, the degenerate FWM process $2\omega_p \rightarrow \omega_i + \omega_s$ and its conjugate up-conversion counterpart $\omega_i + \omega_s \rightarrow 2\omega_p$ meet the requirements for implementing BNOT, where $\omega_p$, $\omega_s$, and $\omega_i$ denote the angular frequencies of the pump, signal, and idler, respectively.}

Looking forward, a promising direction is to integrate BNOT-style tomography into adaptive control and self-calibrating photonic architectures. Real-time forward and backward measurements, combined with Bayesian~\cite{Higgins2007EntanglementFree,Wiebe2014HamiltonianLearning,Lumino2018ExpPhaseEstimationML,Wang2017ExperimentalQuantumHamiltonianLearning} or machine-learning estimators~\cite{Zheng2024GlobalCalibrationPIC}, may enable automatic compensation of coupling losses across large photonic arrays and continuous tracking of drift or degradation. Such closed-loop implementations have the potential to transform coupling calibration from a post-measurement diagnostic into an active subsystem of photonic hardware, supporting reproducible, system-level benchmarks as nonlinear PICs transition from laboratory demonstrations to scalable quantum and classical technologies.

%
%%%%%%%%%%%%%%%%%%%
\begin{acknowledgments}
The authors thank Avik Dutt, Xinyi Ren, and Sri Krishna Vadlamani for fruitful discussions. This work was supported by funding from the DARPA INSPIRED program.
\end{acknowledgments}
%
%%%%%%%%%%%%%%%%%%%
\appendix

\begin{figure}
    \centering
	{\centering\includegraphics[width=1\linewidth]{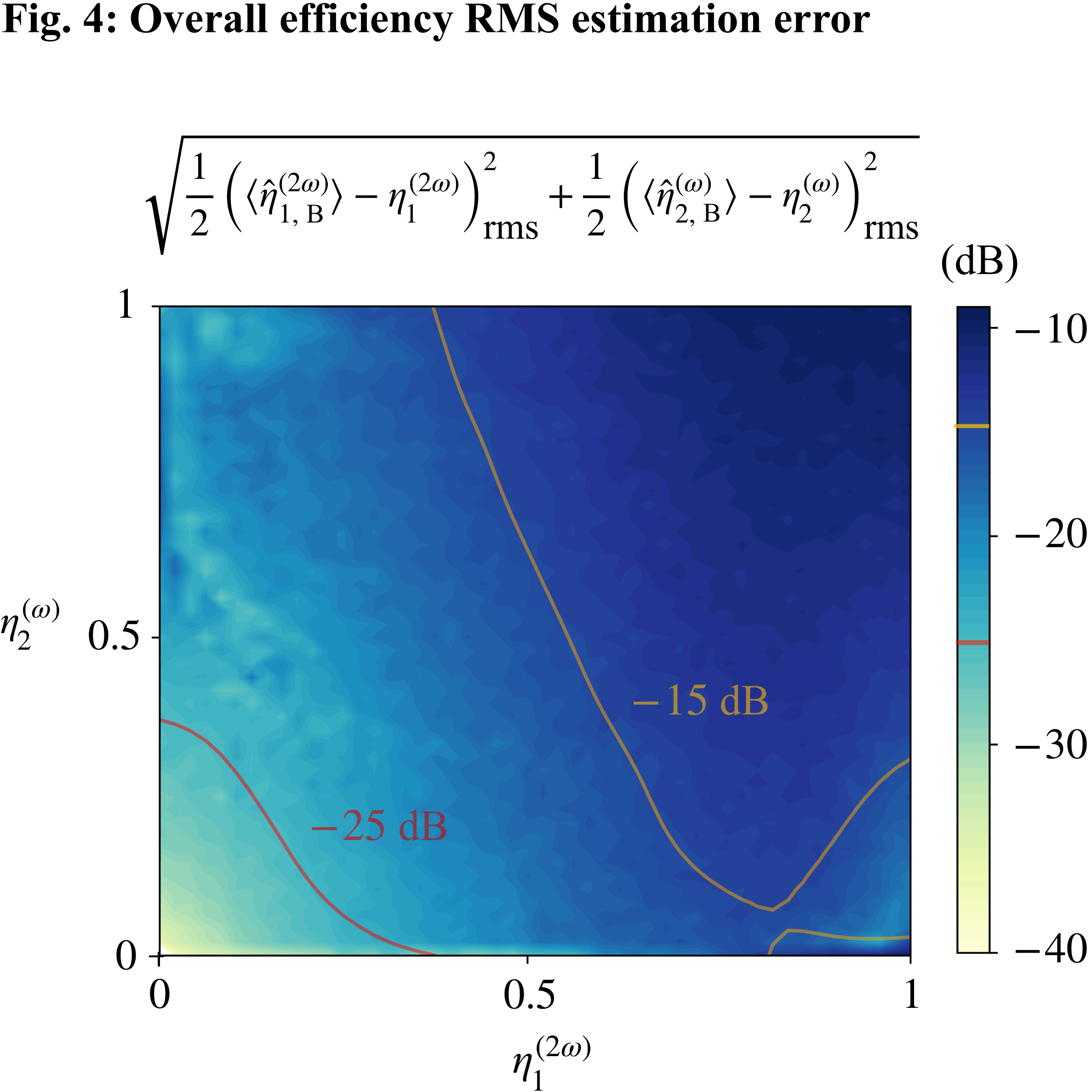}}
	\caption{\BH{The combined RMS errors of the estimated coupling efficiencies $\hat{\eta}^{(2\omega)}_1$ and $\hat{\eta}^{(\omega)}_2$ in Fig.~\ref{fig:MSE} over the full on/off-chip efficiency space $\eta^{(2\omega)}_1, \eta^{(\omega)}_2 \in (0,1)$.}}
	\label{fig:Total_RMSE}
\end{figure}

\BH{\section{Total RMS estimation error in Fig.~\ref{fig:MSE}}}
\label{sec:Total_RMSE}
\BH{To quantify the overall estimation accuracy, we combine the individual RMS errors of $\eta^{(2\omega)}_1$ and $\eta^{(\omega)}_2$ from Fig.~\ref{fig:MSE} into a single metric defined as the root-mean-square of their combined deviations. The resulting heatmap in Fig.~\ref{fig:Total_RMSE} shows the joint estimation error.}
\vspace{5mm}
\section{Analytical formula derivation}
\label{sec:analytical}
In this section, we analytically derive the squeezing and SHG dynamics generated in a PPLN device. Let $\hat{a}_{\omega}$ and $\hat{a}_{2\omega}$ denote the field operators at frequencies $\omega$ and $2\omega$, respectively. Both nonlinear interactions arise from the same three–wave mixing Hamiltonian at PPLN position $z$, given by
\begin{equation}
\hat{\mathcal{H}}
=\frac{\hbar g}{2i}\left(\hat{a}_{\omega}^{2}\,\hat{a}_{2\omega}^{\dagger}\,e^{i\Delta k\,z}
-\hat{a}_{\omega}^{\dagger\,2}\,\hat{a}_{2\omega}\,e^{-i\Delta k\,z}\right),
\end{equation}
where $g$ is the nonlinear coupling strength, $\Delta k = 2k_{\omega}-k_{2\omega}$ is the wave-vector mismatch. 

Thus, the temporal evolution of the operators are governed by the Heisenberg–Langevin (HL) equations
\begin{equation}
\begin{aligned}
        \frac{d\,\hat{a}_{2\omega}}{dz}&=-\frac{g}{2}\,\hat{a}^2_{\omega}\,e^{i\Delta k\,z}-\frac{\Gamma_{2\omega}}{2}\,\hat{a}_{2\omega}+\sqrt{\Gamma_{2\omega}}\,\hat{v}_{2\omega}\\
        \frac{d\,\hat{a}_{\omega}}{dz}&=g\,\hat{a}_\omega^\dagger\,\hat{a}_{2\omega}\,e^{-i\Delta k\,z}-\frac{\Gamma_{\omega}}{2}\,\hat{a}_\omega+\sqrt{\Gamma_{\omega}}\,\hat{v}_\omega,
\end{aligned}
\end{equation}
where $\Gamma_{2\omega}$ and $\Gamma_{\omega}$ are the PPLN waveguide propagation losses at the corresponding modes, $\hat{a}_{2\omega}$ and $\hat{a}_{\omega}$ are the field operators, $\hat{v}_{2\omega}$ and $\hat{v}_{\omega}$ are the vacuum noise operators. The field and vacuum operators satisfy the commutation relations: $\left[\hat{a}_{n},\hat{a}^{\dagger}_{m}\right]=\left[\hat{v}_{n},\hat{v}^{\dagger}_{m}\right]=\delta_{n\mu},\langle\hat{a}^\dagger_{n}(0)\,\hat{a}_{m}(0)\rangle =\langle\hat{v}^{\dagger}_{n}\,\hat{v}_{m}\rangle=0$ for $n,\,m\in\{2\omega,\omega\}$.
Although both squeezing and SHG processes originate from the same interaction Hamiltonian, the relevant approximations and boundary conditions differ. In the following subsections, we solve the Heisenberg–Langevin equation of SHG and squeezing. 

\subsection{Second-harmonic generation}
\subsubsection{Single-pass waveguide field}
In SHG, the pump field at frequency $\omega$ is treated as a strong, undepleted classical drive such that $\langle\hat{a}_{\omega}\rangle\approx A_{\omega}$ remains constant along the waveguide and the non-linear coupling strength of SHG is
\begin{equation}
    g=\frac{2\omega\,d_\text{eff}}{n_{2\omega}\,c}\sqrt{\frac{\hbar\omega\,S_{2\omega}}{\epsilon_0\,S^2_\omega\,L}},
    \label{eq:g_constant}
\end{equation}
where $d_{\text{eff}}=2d_{33}/\pi$ is the effective nonlinear coefficient under quasi-phase-matching, $d_{33}$ is the nonlinear optical coefficient, $c$ is the speed of light, $\epsilon_{0}$ denotes the vacuum permittivity, $n_{2\omega}$ is the refractive index of LiN at $2\omega$, $S_{2\omega}$ and $S_{\omega}$ are the effective mode areas at the corresponding modes, and $L$ is the poling length of PPLN.

Under this approximation, the second-harmonic mode $\hat{a}_{2\omega}$ obeys a driven linear equation, and its mean-field solution at position $z=L$ is
\begin{equation}
\begin{aligned}
        \langle\hat{a}_{2\omega}(L)\rangle&\approx-\frac{g\,A_\omega^2}{2}\,\int_{0}^{L}\,dz'\,e^{\left(i\Delta k+\frac{\Gamma_{2\omega}}{2}\right)\,(z'-L)}\\
        &=-\frac{g\,A_\omega^2}{2}\,\left(\frac{1-e^{-\left(i\Delta k+\frac{\Gamma_{2\omega}}{2}\right)\,L}}{i\Delta k+\Gamma_{2\omega}/2}\right).
\end{aligned}
\end{equation}
where $\Delta k$ is the wave-vector mismatch, and $\Gamma_{2\omega}$ is the loss of the generated second-harmonic field.
\begin{equation}
        |\langle\hat{a}_{2\omega} (L)\rangle|^2=g^2\,A_\omega^4\,\left(\frac{1+e^{-\Gamma_{2\omega}L}-2e^{-\frac{\Gamma_{2\omega}L}{2}}\cos{\left(\Delta k\,L\right)}}{4\Delta k^2+\Gamma^2_{2\omega}}\right).
\end{equation}

\BH{\subsubsection{Fabry–Pérot effects from the waveguide facets}
\label{sec:FP_SHG}
When reflections occur at the waveguide facets, the pump field forms a weak Fabry–Pérot (FP) cavity inside the bus waveguide. Under resonant constructive interference, the reflected fields generate circulating pump components that modifies the intracavity pump amplitude $A_{\omega}^2$ as
\begin{equation}
A_{\omega}^2\rightarrow\frac{A_{\omega}^2}{\left(1-|R_{\omega}|\right)^2+4|R_{\omega}|\,\sin^2{\left(n_{\omega}k_{\omega}L\right)}},
\label{eq:FP_SHG}
\end{equation}
i.e., $k_\omega=\omega/c$, which in turn leads to the second-harmonic field intensity
\begin{equation}
|\langle\hat{a}_{\text{ON}}^{(2\omega)}\rangle
|^2=\frac{|\langle\hat{a}_{2\omega} (L)\rangle|^2}{\left(\left(1-|R_{\omega}|\right)^2+4|R_{\omega}|\,\sin^2{\left(n_{\omega}k_{\omega}L\right)}\right)^2}.
\label{eq:FP_SHG}
\end{equation}
Here, $|R_{\omega}|$ denotes the reflectivity of each bus-waveguide facet at $\omega$.}

\subsubsection{On-chip SHG efficiency}
In this experiment, the second-harmonic field is generated inside the waveguide and is therefore described by its on-chip electric-field amplitude. In contrast, the pump field is specified by the optical power delivered to the chip through the on/off chip coupling efficiency $\eta_2^{(\omega)}$. To evaluate the detected SHG efficiency, we convert field amplitudes to optical powers using
\begin{equation}
P_{\Omega}=\frac{n_{\Omega}\,c\,\epsilon_{0}}{2}\,|E^{(\Omega)}|^{2}\,S_{\Omega},
\label{eq:power}
\end{equation}
where $\Omega\in\{\omega,2\omega\}$, $n_{\Omega}$ is the refractive index at frequency $\Omega$, $S_{\Omega}$ denotes the effective mode area, $c$ is the speed of light, and $\epsilon_{0}$ is the vacuum permittivity.

To evaluate the optical powers using Eq.~\eqref{eq:power}, we relate the electric-field amplitudes to the corresponding quantum-field variables. The second-harmonic field is generated inside the waveguide and is therefore described by its on-chip electric-field amplitude. In contrast, the pump field is specified by the off-chip input power and couples into the waveguide with efficiency $x_2$. Accordingly, the electric-field amplitudes are written as
\begin{equation}
\begin{aligned}
    E_{\text{ON}}^{(2\omega)}&=\frac{\langle\hat{a}_{\text{ON}}^{(2\omega)}\rangle}{n_{2\omega}}\,\sqrt{\frac{4\hbar\omega}{\epsilon_0\,S_{2\omega} L}}\;\;,\;\;
    E_{\text{OFF}}^{(\omega)}=\frac{A_{\omega}}{n_\omega}\sqrt{\frac{2\hbar\omega}{x_2\,\epsilon_0\,S_\omega L}},
    \label{eq:E_field}
\end{aligned}
\end{equation}
The last factors refer to the single-photon electric-field amplitudes. Consequently, $|\langle\hat{a}_{\text{ON}}^{(2\omega)}\rangle|^2$ and $|A_\omega|^2$ represent the on-chip mean photon numbers at the corresponding frequencies, while the fields above provide the physical electric-field amplitudes required to evaluate the optical powers through Eq.~\eqref{eq:power}.

Using this relation, the on-chip second-harmonic power $P_{\text{ON}}^{(2\omega)}$ referenced to the off-chip pump power $P_\omega^{(\text{OFF})}$ leads to the on-chip SHG efficiency~\cite{yang2024symmetric}:
\begin{equation}
    \begin{aligned}
        \tilde{\mathcal{E}}_\text{ON}(x_2)&\equiv\frac{\varepsilon^2_2\,P_{\text{ON}}^{(2\omega)}}{\left(P_\omega^{(\text{OFF})}\,L\right)^2}=\frac{2\,n_{2\omega}\,S_{2\omega}}{n^2_{\omega}\,\epsilon_0\,c\,L^2\,S^2_{\omega}}\,\varepsilon^2_2\,\left|\frac{E_{2\omega}}{E_{\omega}^{(\text{OFF})\,2}}\right|^2\\
        &=\beta\,x_2^2\,\varepsilon^2_2\,\left(\frac{1+e^{-\Gamma_{2\omega}L}-2e^{-\frac{\Gamma_{2\omega}L}{2}}\cos{\left(\Delta k\,L\right)}}{\Delta k^2+\Gamma^2_{2\omega}/4}\right),
        \label{eq:on_ship_SHG}
    \end{aligned}
\end{equation}
where $\varepsilon_2\sim \mathcal{N}(1,\sigma^2_{\omega})$ denotes the off-chip pump power fluctuation rate and
\begin{equation}
\beta=\frac{2\,k_{\omega}^2\,d^2_\text{eff}\,S_{2\omega}}{n_{2\omega}\,n_{\omega}^2\,c\,\epsilon_0\,L^2\,S_\omega^2}\,\frac{1}{\left(\left(1-|R_{\omega}|\right)^2+4|R_{\omega}|\,\sin^2{\left(n_{\omega}k_{\omega}L\right)}\right)^2}.
\end{equation}
\BH{From Eq.~\eqref{eq:on_ship_SHG}, the ideal on-chip SHG efficiency is defined as
\begin{equation}
    \mathcal{E}_\text{ON}(\eta_2^{(\omega)})\equiv 
    \lim_{ \substack{\varepsilon_2 \rightarrow 1 \\ \Delta k \rightarrow 0} }
    \tilde{\mathcal{E}}_\text{ON}(\eta_2^{(\omega)}),
\end{equation}
excluding all the uncertainties of $\tilde{\mathcal{E}}_\text{ON}$.
}

\subsection{Squeezed light}
\BH{\subsubsection{Single-pass waveguide field}
Squeezed-light generation in the PPLN waveguide arises from the SPDC process in which the $2\omega$ pump is converted into photon pairs at $\omega$. Under strong pumping, the $2\omega$ mode is treated as a classical undepleted drive, $\langle\hat{a}_{2\omega}\rangle\approx A_{2\omega}\in\mathbb{R}$, so the HL equations of $\hat{a}_\omega$ and $\hat{a}^\dagger_\omega$ follow:
\begin{equation}
\begin{aligned}
    \frac{d\,\hat{a}_{\omega}}{dz}&=\mu\,e^{-i\Delta k \,z}\,\hat{a}^{\dagger}_{\omega}-\frac{\Gamma_{\omega}}{2}\,\hat{a}_\omega+\sqrt{\Gamma_{\omega}}\,\hat{v}_\omega,\\
    \frac{d\,\hat{a}^\dagger_{\omega}}{dz}&=\mu\,e^{i\Delta k \,z}\,\hat{a}_{\omega}-\frac{\Gamma_{\omega}}{2}\,\hat{a}^{\dagger}_\omega+\sqrt{\Gamma_{\omega}}\,\hat{v}^{\dagger}_\omega
\end{aligned}
    \label{eq:HL_eq_sqz}
\end{equation}
with $\mu\equiv g\,A_{2\omega}$. Moving to the rotating frame: $\hat{a}_\omega=\hat{b}_\omega\,e^{-i\Delta k\,z/2}$ and $\hat{v}_\omega=\hat{\tilde{v}}_\omega\,e^{-i\Delta k\,z/2}$, Eq.~\eqref{eq:HL_eq_sqz} becomes
\begin{equation}
\begin{aligned}
        \frac{d\,\hat{b}_{\omega}}{dz}&=\mu\,\hat{b}^{\dagger}_{\omega}-\left(\frac{\Gamma_{\omega}}{2}-i\frac{\Delta k}{2}\right)\,\hat{b}_\omega+\sqrt{\Gamma_{\omega}}\,\hat{\tilde{v}}_\omega,\\
        \frac{d\,\hat{b}^{\dagger}_{\omega}}{dz}&=\mu\,\hat{b}_{\omega}-\left(\frac{\Gamma_{\omega}}{2}+i\frac{\Delta k}{2}\right)\,\hat{b}^{\dagger}_\omega+\sqrt{\Gamma_{\omega}}\,\hat{\tilde{v}}^{\dagger}_\omega.
\end{aligned}
    \label{eq:HL_eq_b_sqz}
\end{equation}
Therefore, the solution of Eq.~\eqref{eq:HL_eq_b_sqz} at $z=L$ is
\begin{widetext}
    \begin{equation}
\begin{aligned}
    \hat{b}_\omega(L)=\frac{2\sqrt{\Gamma_{\omega}}\,\left(\left(\Gamma_{\omega}+i\Delta k\right)\,\hat{\tilde{v}}_\omega+2\mu\,\hat{\tilde{v}}_\omega^\dagger\right)}{\Gamma_{\omega}^2-\Delta K^2}+e^{-\frac{\Gamma_{\omega}}{2}\,L}\,\hat{b}_\omega(0)\,\cosh{\left(\frac{\Delta K}{2}\,L\right)}+e^{-\frac{\Gamma_{\omega}}{2}\,L}\,\left(\frac{i\Delta k\,\hat{b}_\omega(0)+2\mu\,\hat{b}^\dagger_\omega(0)}{\Delta K}\right)\,\sinh{\left(\frac{\Delta K}{2}\,L\right)},
\end{aligned}
\end{equation}
\end{widetext}
where $\Delta K=\sqrt{4\mu^2-\Delta k^2}$. Based on the quadrature convention $\hat{q}_\omega\equiv\hat{b}_\omega+\hat{b}_\omega^\dagger$ and $\hat{p}_\omega\equiv(\hat{b}_\omega-\hat{b}_\omega^\dagger)/i$, we obtain $\hat{b}_\omega$' quadrature variances and the covariance sum:
\begin{widetext}
    \begin{equation}
\begin{aligned}
        \Delta q^2_\omega(\mu;\Delta k)&=\frac{4\Gamma_{\omega}\left(\Delta k^2+\left(\Gamma_{\omega}+2\mu\right)^2\right)}{\left(\Gamma_{\omega}^2-\Delta K^2\right)^2}-\frac{e^{-\Gamma_{\omega}\,L}}{\Delta K^2}\left\{\Delta k^2-2\,\mu\,\left(2\mu\,\cosh{\left(\Delta K\,L\right)}+\Delta K\,\sinh{\left(\Delta K\,L\right)}\right)\right\}\quad\quad(\text{anti-squeezing}),\\
        \Delta p^2_\omega(\mu;\Delta k)&=\frac{4\Gamma_{\omega}\left(\Delta k^2+\left(\Gamma_{\omega}-2\mu\right)^2\right)}{\left(\Gamma^2-\Delta K^2\right)^2}-\frac{e^{-\Gamma_{\omega}\,L}}{\Delta K^2}\left\{\Delta k^2-2\,\mu\,\left(2\mu\,\cosh{\left(\Delta K\,L\right)}-\Delta K\,\sinh{\left(\Delta K\,L\right)}\right)\right\}\quad\quad(\text{squeezing}),\\
        \mathcal{C}_\omega(\mu;\Delta k)&\equiv\text{Cov}\left(\hat{q}_\omega(\mu;\Delta k),\hat{p}_\omega(\mu;\Delta k)\right)+\text{Cov}\left(\hat{p}_\omega(\mu;\Delta k),\hat{q}_\omega(\mu;\Delta k)\right)=8\,\Delta k\,\mu\left\{\frac{4\,\Gamma_{\omega}}{\left(\Gamma_{\omega}^2-\Delta K^2\right)^2}+\frac{e^{-\Gamma_{\omega}\,L}}{\Delta K^2}\,\sinh{\left(\frac{\Delta K}{2}\,L\right)}\right\},
        \label{eq:Variance}
\end{aligned}
\end{equation}
\end{widetext}
where $\text{Cov}(\hat{A},\hat{B})\equiv\langle\hat{A}\,\hat{B}\rangle-\langle\hat{A}\rangle\,\langle\hat{B}\rangle$ denotes the covariance between operators $\hat{A}$ and $\hat{B}$.}

\BH{\subsubsection{Fabry–Pérot effects from the waveguide facets}
Similar to the SHG case discussed in Appendix~\ref{sec:FP_SHG}, facet reflections form a weak FP cavity that modifies the pump field at frequency $2\omega$ driving the squeezing process.  As a result, the nonlinear drive $\mu$ is rescaled by the corresponding FP factor,
\begin{equation}
\begin{aligned}
        \mu\rightarrow\mu_\text{eff}=\frac{\mu}{\sqrt{\left(1-|R_{2\omega}|\right)^2+4|R_{2\omega}|\,\sin^2{\left(n_{2\omega}k_{2\omega}L\right)}}},
\end{aligned}
\end{equation}
i.e., $k_{2\omega}=2\omega/c$, where $|R_{2\omega}|$ denotes the facet reflectivity at $2\omega$.}

\BH{\subsubsection{On-chip squeezing levels}
The field quadrature at $z=L$ is read out using balanced homodyne detection, where $\hat{b}_\omega(L)$ interferes with a local oscillator $\sqrt{N_\text{LO}}\,e^{i(\theta+\xi)}$ on a 50:50 beamsplitter, followed by photodetection and subtraction of the photocurrents. Here, $\theta$ denotes the stabilized mean phase of the local oscillator, which sets the measured quadrature. The phase fluctuation $\xi\sim\mathcal{N}(0,\sigma^2_\text{PN})$ models the LO phase noise arising from residual path-length fluctuations between the two interferometer arms, where $\sigma_\text{PN}$ characterizes the standard deviation of the phase noise. These fluctuations rotate the measurement basis and mix the squeezed and anti-squeezed quadratures, as described by
    \begin{equation}
\begin{aligned}
        \Delta Q^2_{\omega,\,\theta+\xi}(\mu_\text{eff};\Delta k)&=\Delta q^{2}_\omega\left(\mu_\text{eff};\Delta k\right)\,\cos^2(\theta+\xi)\\
        &+\Delta p^{2}_\omega\left(\mu_\text{eff};\Delta k\right)\,\sin^2(\theta+\xi)\\
        &+\,\mathcal{C}_\omega\left(\mu_\text{eff};\Delta k\right)\,\sin(\theta+\xi)\,\cos(\theta+\xi).
        \label{eq:Delta_Q}
\end{aligned}
\end{equation}}
\begin{figure}[h]
    \centering
	{\centering\includegraphics[width=0.87\linewidth]{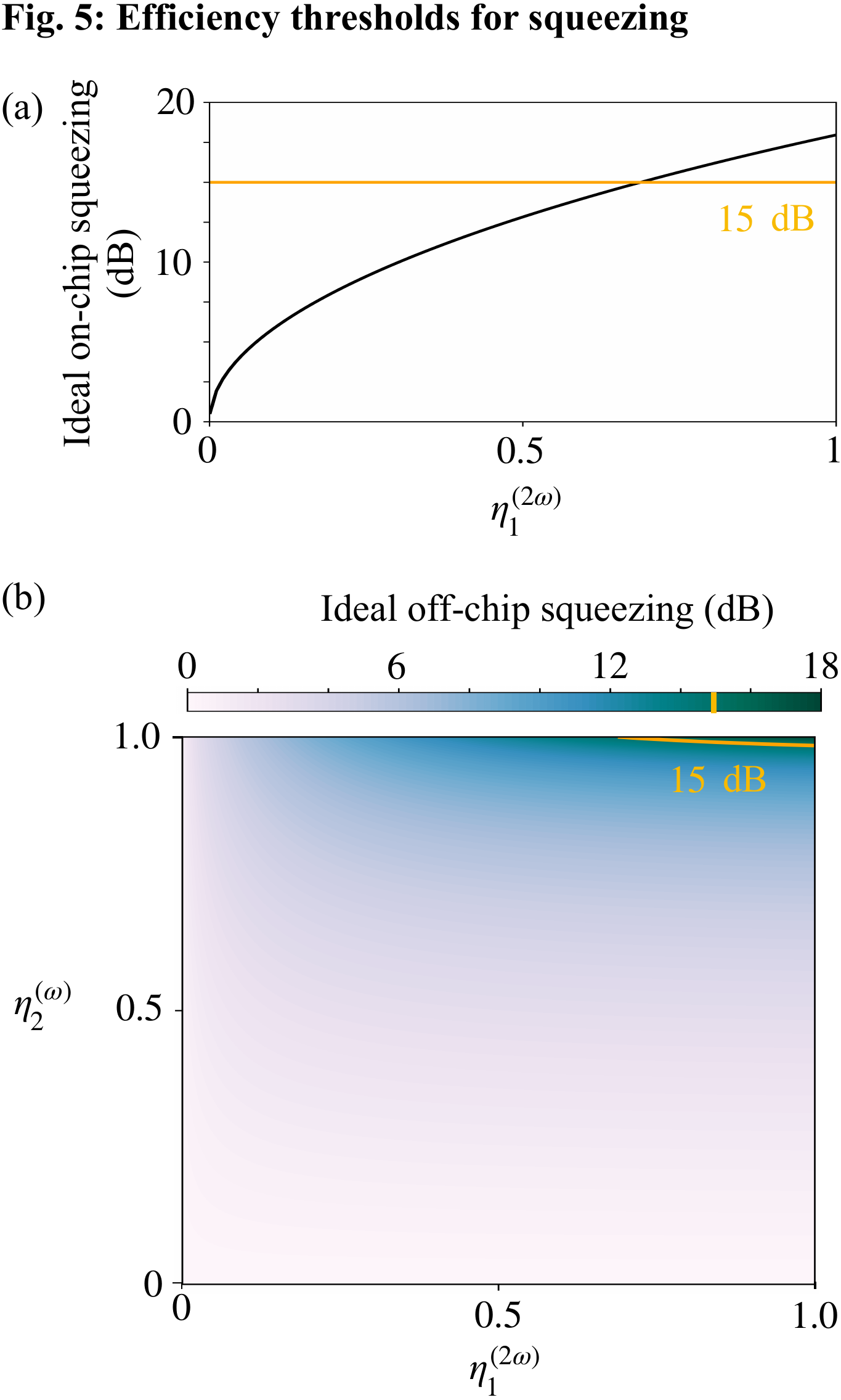}}
	\caption{(a) \BH{Ideal} on-chip squeezing versus $\eta_1^{(2\omega)}$. (b) \BH{Ideal} off-chip squeezing versus $\eta_1^{(2\omega)}$ and $\eta_2^{(\omega)}$. The orange horizontal line in (a) and curve in (b) denote the on-chip and off-chip $15$-dB squeezing threshold.}
	\label{fig:Sqz_15dB}
\end{figure}
\BH{In this work, the on-chip squeezing and anti-squeezing levels are defined as the squeezing and anti-squeezing of the optical field generated inside the device, before any output coupling loss. They are given by
\begin{equation}
    \begin{aligned}
        \tilde{\mathcal{S}}_\text{ON}(x_1)
        &= -10\,\log_{10}\left[\Delta Q_{\omega,\, \pi/2+\xi}^2\bigl(\alpha\sqrt{x_1\varepsilon_1};\,\Delta k\bigr)\right],\\
        \tilde{\mathcal{A}}_\text{ON}(x_1)
        &= 10\,\log_{10}\left[\Delta Q_{\omega,\,\xi}^2\bigl(\alpha\sqrt{x_1\varepsilon_1};\,\Delta k\bigr)\right],
        \label{eq:on_chip_sqz}
    \end{aligned}
\end{equation}
where
\begin{equation}
\alpha=g\,\sqrt{\frac{n_{2\omega}\,L\,P_{2\omega}}{2\hbar\omega c}\,\frac{1}{\left(1-|R_{2\omega}|\right)^2+4|R_{2\omega}|\,\sin^2{\left(n_{2\omega}k_{2\omega}L\right)}}}.
\label{eq:alpha}
\end{equation}
To obtain Eq.~\eqref{eq:alpha}, we relate the effective nonlinear drive,
\begin{equation}
    \begin{aligned}
        A_{2\omega}&=\sqrt{\frac{2x_1\,\varepsilon_1\,P_{2\omega}}{n_{2\omega}\,c\,\epsilon_{0}\,S_{2\omega}}\left(\frac{n_{2\omega}^2\,\epsilon_0\,S_{2\omega}\,L}{4\,\hbar\,\omega}\right)}=\sqrt{\frac{n_{2\omega}\,L\,x_1\,\varepsilon_1\,P_{2\omega}}{2\,\hbar\,\omega\,c}},
    \end{aligned}
\end{equation}
to the off-chip pump power, $P_{2\omega}$, via Eq.~\eqref{eq:power} and substitute this expression into $\mu_\text{eff}=g\,A_{2\omega}$ to yield
\begin{equation}
\mu_\text{eff} = \alpha\,\sqrt{x_1\,\varepsilon_1}.
\end{equation}
Specifically, $x_1$ is the tunable coupling efficiency, and $\varepsilon_1\sim \mathcal{N}(1,\sigma^2_{2\omega})$ characterizes the STD of the pump power fluctuation rate.}

\BH{From Eq.~\eqref{eq:on_chip_sqz}, the ideal on-chip squeezing level is thus defined as
\begin{equation}
\begin{aligned}
        \mathcal{S}_\text{ON}(\eta_1^{(2\omega)})&\equiv 
    \lim_{ \substack{\varepsilon_1 \rightarrow 1 \\ \Delta k \rightarrow 0\\ \xi \rightarrow 0} }
    \tilde{\mathcal{S}}_\text{ON}(\eta_1^{(2\omega)}),
\end{aligned}
\end{equation}
excluding all the uncertainties of $\tilde{\mathcal{S}}_\text{ON}$.
}
%%%%%%%%%%%%%%%%%%%
\vspace{5mm}
\section{Efficiency thresholds for off-chip squeezing}
\label{sec:threshold}
Fig.~\ref{fig:Sqz_15dB} summarizes the dependence of \BH{ideal} on- and off-chip squeezing on \BH{on/off-coupling} efficiencies. Fig.~\ref{fig:Sqz_15dB}(a) shows the on-chip squeezing $\mathcal{S}_\text{ON}$ as a function of the coupling efficiency $\eta^{(2\omega)}_{1}$, while Fig.~\ref{fig:Sqz_15dB}(b) maps the off-chip squeezing $\mathcal{S}_\text{OFF}$ as a joint function of $\eta^{(2\omega)}_{1}$ and $\eta^{(\omega)}_{2}$. The orange lines in both panels mark the squeezing contour $15~\text{dB}$, a benchmark frequently cited in advanced quantum technologies, such as CV quantum error correction and precision quantum metrology~\cite{fukui2018high,larsen2025nature}.

Achieving this on-chip benchmark requires an on/off chip coupling efficiency $\eta^{(2\omega)}_{1} \gtrsim 0.65$, which is attainable using current photonic integration technologies. For example, edge-coupled LiN modulators have demonstrated per-coupling efficiencies below $0.5~\text{dB}$ (corresponding to efficiencies $>90~\%$) at $1550~\text{nm}$~\cite{Ying2021LowLossEdge}, indicating that such coupling efficiencies are experimentally feasible. However, maintaining this performance off chip imposes stricter requirements. For example, to retain the off-chip squeezing level $\mathcal{S}_\text{OFF}\gtrsim 15~\text{dB}$, the output \BH{on/off-coupling} efficiency must satisfy $\eta^{(\omega)}_{2}\gtrsim 0.97$.

\bibliographystyle{apsrev4-2}
\bibliography{references}

\end{document}